\documentclass[]{interact}

\usepackage{graphicx}
\usepackage{accsupp}

\usepackage[bookmarks=false,hidelinks]{hyperref}

\usepackage{epstopdf}
\usepackage[caption=false]{subfig}

\usepackage{float}

\usepackage[natbibapa,nodoi]{apacite}
\makeatletter
\newcommand\footnoteref[1]{\protected@xdef\@thefnmark{\ref{#1}}\@footnotemark}
\makeatother

\theoremstyle{plain}

\theoremstyle{definition}

\theoremstyle{remark}

\usepackage{mathtools}
\usepackage{amssymb}

\def\SuF{{\fontfamily{cmtt}\selectfont{SuF}}}
\def\SpF{{\fontfamily{cmtt}\selectfont{SpF}}}
\def\CoF{{\fontfamily{cmtt}\selectfont{CoF}}}
\def\CNN{{\fontfamily{cmtt}\selectfont{CNN}}}
\def\RNN{{\fontfamily{cmtt}\selectfont{RNN}}}

\def\normal{{\fontfamily{cmtt}\selectfont{NR}}}
\def\deadpan{{\fontfamily{cmtt}\selectfont{DP}}}
\def\speed{{\fontfamily{cmtt}\selectfont{SP}}}
\def\nr12{{\fontfamily{cmtt}\selectfont{NR12}}}
\def\vc12{{\fontfamily{cmtt}\selectfont{VC-NR12}}}
\def\and{{\fontfamily{cmtt}\selectfont{and}}}

\begin{document}

\title{Annotation of Soft Onsets in String Ensemble Recordings}

\author{
\name{Maciej Tomczak,\textsuperscript{a}\thanks{CONTACT Massimiliano Di Luca. Email: m.diluca@bham.ac.uk}
Min Susan Li,\textsuperscript{b} 
Adrian Bradbury,\textsuperscript{b} 
Mark Elliott,\textsuperscript{c,f} 
Ryan Stables,\textsuperscript{a} 
Maria Witek,\textsuperscript{d}  
Tom Goodman,\textsuperscript{c} 
Diar Abdlkarim,\textsuperscript{b} 
Massimiliano Di Luca,\textsuperscript{a} 
Alan Wing,\textsuperscript{e} and 
Jason Hockman\textsuperscript{a,g}
}
\affil{
\textsuperscript{a}Sound and Music Analysis (SoMA) Group in Digital Media Technology (DMT) Lab, Birmingham City University, UK;\\
\textsuperscript{b}VR Lab, School of Psychology, University of Birmingham, UK;\\
\textsuperscript{c}WMG, University of Warwick, Coventry, UK;\\
\textsuperscript{d}Department of Music, University of Birmingham, UK;\\
\textsuperscript{e}Active Touch Lab, School of Psychology, University of Birmingham, UK;\\
\textsuperscript{f}School of Sport, Exercise and Rehabilitation Sciences, University of Birmingham, UK;\\
\textsuperscript{g}School of Digital Arts (SODA), Manchester Metropolitan University, Manchester, UK}}
\maketitle

\begin{abstract}
Onset detection is the process of identifying the start points of musical note events within an audio recording. While the detection of percussive onsets is often considered a solved problem, \textit{soft} onsets---as found in string instrument recordings---still pose a significant challenge for state-of-the-art algorithms. The problem is further exacerbated by a paucity of data containing expert annotations and research related to best practices for curating soft onset annotations for string instruments. To this end, we investigate inter-annotator agreement between 24 participants, extend an algorithm for determining the most consistent annotator, and compare the performance of human annotators and five established onset detection algorithms. Experimental results reveal a positive relationship between musical experience and both inter-annotator agreement and performance in comparison with automated systems. Additionally, onsets produced by fingered pitch transitions, as well as those from the cello recordings, were found to be particularly challenging for both human annotators and automatic approaches. To promote further research, we publicly release all experimental data associated with this study, alongside the \textit{Virtuoso Strings} Haydn subset, consisting of multitrack recordings of the Op.~74 No.~1 Finale performed under varying expressive conditions, each with corresponding individual instrumental onset annotations.
\end{abstract}


\section{Introduction}\label{sec:intro}
Within the field of music information retrieval, note onset detection seeks to identify the start points of musical events in an audio recording. It serves as a fundamental step in applications ranging from automatic transcription and tempo estimation to interactive performance systems used in creative and educational industries. 

The majority of onset detection approaches rely on subsampled versions of the audio signal (i.e., detection functions), which exhibit peaks at time points where specific features of the signal change (e.g., energy, spectral content). Task performance is typically evaluated using standard metrics that assess onset presence within tolerance windows surrounding ground truth annotations. Unlike symbolic representations~(e.g., MIDI) where onsets are defined as discrete events, audio waveforms present a continuous signal in which the precise location of an onset is often ambiguous. Furthermore, variability in strategies adopted by annotators can widen the discrepancy of ground truth labels. 

While onset detection for percussive sounds achieves near-perfect performance in solo contexts \citep{wu2018review}, detecting \textit{soft} onsets remains a non-trivial task. In~this work, we define soft onsets as musical events characterised by a gradual rise in the amplitude envelope (long rise time) and a lack of prominent transient broadband energy, resulting in a sound that emerges gradually rather than abruptly. This lack of distinct acoustic features complicates the manual annotation process; soft onset annotation is inherently more difficult than percussive annotation and typically yields lower inter-annotator agreement. Consequently, establishing reliable ground truth for soft onsets poses a significant challenge for both music perception research and algorithm development.

Existing research has produced freely available datasets\footnote{\url{https://www.ismir.net/resources/datasets/}} of soft onset annotations, such as EEP~(Marchini et al., \citeyear{marchini2014sense}),\footnote{\url{https://upf.edu/web/mtg/eep}} URMP~(Li et al., \citeyear{li2018creating}),\footnote{\url{https://labsites.rochester.edu/air/projects/URMP.html}} and CMMSD~\citep{von2014cmmsd}.\footnote{\url{https://sourceforge.net/projects/segmentationgt/}} However, the labels in these collections are typically generated from the responses of only a few participants. Furthermore, relatively few studies have systematically investigated annotator strategies, quantified variability of annotations, or highlighted specifically difficult contexts and instruments. 

To this end, we conduct three experiments to analyse annotation variability and introduce two contributions: the \textit{Virtuoso Strings} dataset (a collection of professional string quartet recordings) and a corresponding Haydn annotation dataset featuring contributions from 24 participants. We also extend a pre-existing strategy \citep{leveau2004methodology} for identifying the most consistent annotator to produce reliable ground truth labels. The aim of these experiments is to investigate best practices for consolidating annotations from individuals with varying levels of experience and to benchmark human performance against automatic onset detection approaches. Additionally, our experimental framework and the proposed dataset are well-suited for broader research in computational sound analysis, as well as music perception and cognition.

\subsection{Background}

\subsubsection{Onset Types}\label{sec:note_types}
An important consideration when annotating note onsets in string instrument recordings is distinguishing between production methods (e.g., \textit{bowed}, \textit{fingered}, \textit{plucked}) and understanding their impact on how a listener perceives the start of an event. String instruments typically produce notes with slow attack times, causing the perceptual start of a sound (the perceptual centre or P-centre) to differ from the physical onset of the signal. This phenomenon has been extensively investigated in rhythm perception research (Morton et al., \citeyear{morton1976perceptual}; Danielsen et al., \citeyear{danielsen2019beat}). Consequently, annotators often fail to converge on a single exact timepoint, suggesting that onset position is more accurately modelled as a temporal interval rather than a fixed point. This interval can be visualised as a distribution where the peak represents the most probable onset position and the width indicates the level of uncertainty. 

\begin{figure}[!t]
    \centering
    \BeginAccSupp{method=plain, 
    Alt={Audio waveform with orange vertical lines denoting human note onset annotations with the standard attack, decay, sustain and release envelopes plotted underneath.}}
    \includegraphics[width=\columnwidth]{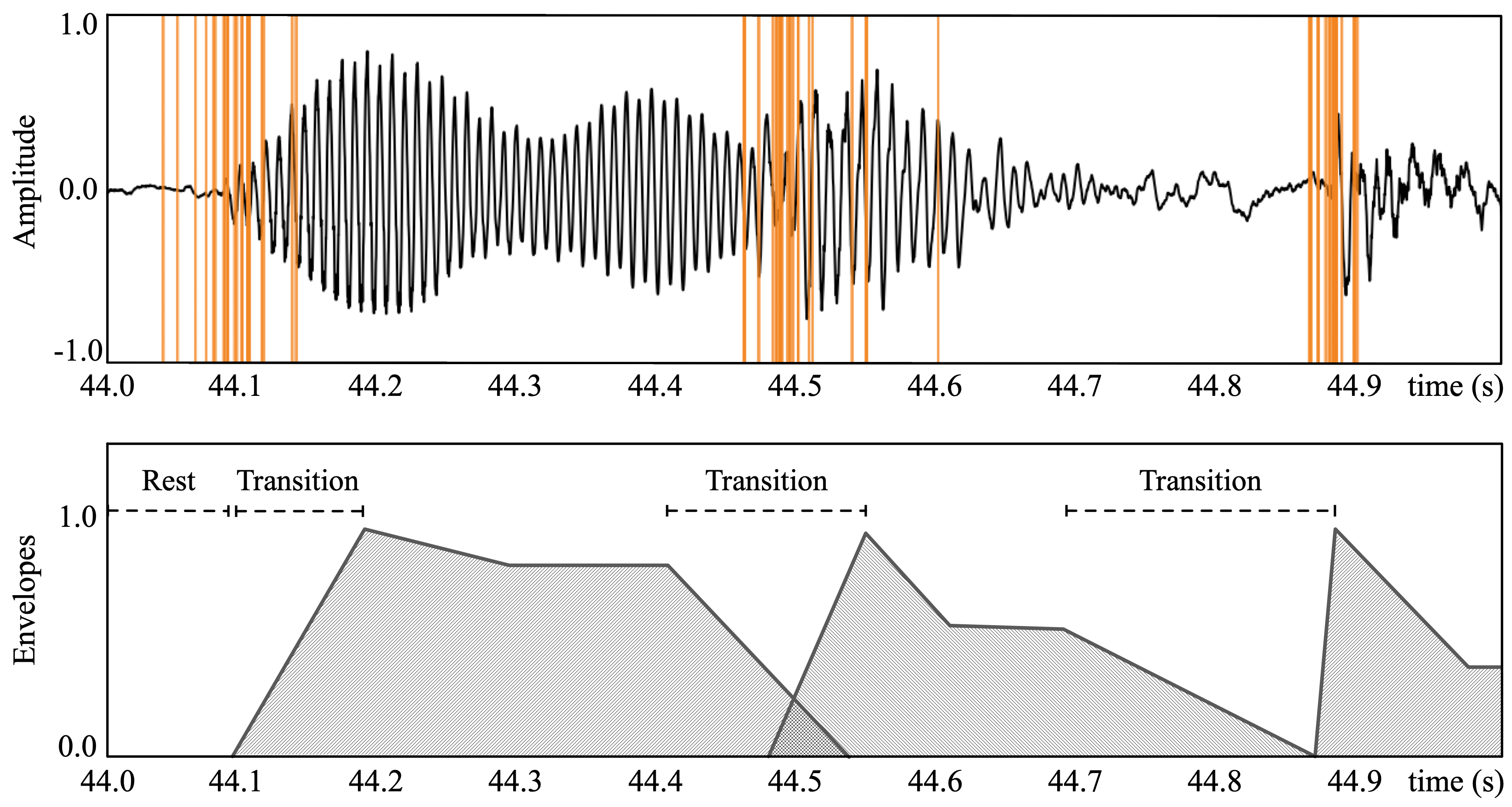}
    \caption{Audio waveform (top) with onset annotations from twenty-four annotators (orange) for a one-second cello excerpt with three notes and envelopes (bottom) with labelled transition and rest regions. The idealised envelopes illustrate attack, decay, sustain and release regions containing annotated note onsets.}
    \label{fig:wav-annotations}
\EndAccSupp{}
\end{figure}

Figure~\ref{fig:wav-annotations} illustrates this variability; it is evident that such annotation data offers valuable insights into human perception. Notably, certain annotations appear slightly ahead of the physical note attack. This can be attributed to factors such as musical anticipation, auditory illusions in rhythm perception, or human error \citep{huron2008sweet,london2012hearing}. While sound onsets can be categorised based on physical or perceptual characteristics \citep{collins2007towards,london2012hearing}, our goal is to associate ground truth annotations of physical onsets with the specific playing techniques of string instruments. To this end, we first introduce a taxonomy based on the physical characterisation of note onsets before defining the specific types investigated in this study.

One of the most fundamental methods for characterisation of music signals relies on envelopes. \Citet{von2014cmmsd} proposed categorising string notes based on standard attack, decay, sustain, and release (ADSR) sections. These sections demarcate note, transition, and rest regions within the audio signal. The bottom panel of Figure~\ref{fig:wav-annotations} depicts idealised ADSR envelopes with highlighted transition and rest regions in a cello performance. This model can be extended to characterise various articulation types. For instance, overlapping note envelopes indicate \textit{legato} transitions (e.g., the first and second notes in Figure~\ref{fig:wav-annotations}). Conversely, \textit{detached} transitions (e.g., the third note in Figure~\ref{fig:wav-annotations}) are represented by note pairs where the release of the first envelope ends immediately before, or coincident with, the attack of the subsequent note. Rest-separated transitions involve musically relevant silence, such as decaying notes that fade gradually without a clear sustaining phase. 

Beyond signal characteristics, understanding the musical context of these changes is crucial. Bowing techniques such as \textit{d\'etach\'e} (separate bow strokes), \textit{martel\'e} (hammered strokes), and \textit{staccato} (short, distinct notes in a single stroke) have been studied in acoustics and mechanics \citep{schoonderwaldt2009mechanics}, as well as in automatic bowing gesture detection \citep{dalmazzo2019bowing}. Recently, \citet{kruger2020playing} implemented a multiclass bag-of-features approach to transcribe eleven string quartet playing techniques, while \citet{falk2023automatic} proposed a convolutional neural network system for transcribing legato violin performances.

We propose three main classes of note onsets for string instruments: \textit{bowed onsets}, characterised by an attack of the bow on a string; \textit{fingered onsets}, characterised by a change in pitch due to the engagement or release of a finger on the fingerboard~(typically the left hand); and \textit{plucked onsets}, due to a finger (typically of the right hand) plucking a string (e.g., pizzicato) \citep{fletcher2012physics}. 

Within each of these classes reside distinct types of onsets. For example, a \textit{bowed onset} might be caused by the bow engaging with the string from above (e.g., spiccato, sautill\'e) \citep{guettler1992bowed}, a change of bowing direction (e.g., up to down bow), a change in bowing pressure or speed within a bow stroke (e.g., portato) \citep{schoonderwaldt2009extraction}, or by a new string being engaged within a bow stroke (e.g., moving from the A to D string). The change in pitch resulting from a \textit{fingered onset} might be due to a new finger engaging with the string (e.g., fourth finger following on from a first finger), or a new finger's engagement being revealed by the release of another finger (e.g., a first finger following on from a third finger); either of these \textit{fingered onset} types (i.e., engagement and release) can also be associated with changes to and from an \textit{open} string on which no fingers are engaged \citep{maestre2009modeling}. \textit{Fingered onsets}, especially those not involving changes to and from open strings, represent a more subtle class of onsets which are better described through pitch information changes rather than the discernible attack regions present in bowed onsets \citep{bello2005tutorial,collins2007towards}.

\subsubsection{Onset Detection}
The task of automatic onset detection based on methods which rely on digital signal processing techniques involves three stages: \textit{preprocessing}, \textit{reduction}, and \textit{peak-picking}~\citep{bello2005tutorial}. The \textit{preprocessing} stage suppresses various aspects of the original audio signal (e.g., specific frequency bands) and enhances the target sounds of the studied musical domain (e.g., accentuating transients of \textit{hard} onsets in drum recordings). The \textit{reduction} stage generates an onset detection function referred to as an activation function in deep learning systems. The detection function outlines occurrences of transients in the form of a subsampled representation of the original signal. The final \textit{peak-picking} stage identifies and selects peaks from the detection or activation function which represent musical events in the original signal. The peak-picking process can be difficult as it relies heavily on clean detection or activation function representations, which can be easily affected when several instrument sounds overlap. A common problem linked to the preprocessing stage is the introduction of additional artefacts that can have a negative effect on the onset detection performance \citep{gillet2008transcription}. As such, current state-of-the-art deep learning onset detection approaches often bypass an explicit preprocessing stage, effectively streamlining the workflow. These systems reduce input representations (such as logarithmic or Mel spectrograms) directly into learned features during system optimisation followed by peak picking on the output activation functions (B{\"o}ck et al., \citeyear{bock2012evaluating}; Schl{\"u}ter \& B{\"o}ck, \citeyear{schluter2014improved}; Ramos et al., \citeyear{ramos2025exploiting}). More recently, systems for audio event detection have incorporated architectures such as convolutional recurrent neural networks (CRNN) and temporal convolutional networks (TCN) (Vogl et al., \citeyear{vogl2017drum}; B{\"o}ck \& Davies, \citeyear{bock2020deconstruct}; Tomczak \& Hockman, \citeyear{tomczak2023onset}).

The MIR evaluation exchange (MIREX) audio onset detection evaluation challenge~\citep{mirex2021onset} is used to test and benchmark new onset detection algorithms on a variety of musical audio signals. 
The ground truth information used in the challenge consists of multiple independent annotations per file (typically three to five, depending on audio complexity). Consequently, there is no single correct reference; instead, the algorithm's mean precision and recall rates are calculated by averaging the scores obtained against each individual annotation. Pinto et al. (\citeyear{pinto2020shift}) proposed a calculation of annotation efficiency based on counting the number of shifts, insertions and deletions. It displays the minimum set of operations needed to modify initial detections for optimal F-measure (see Equation \eqref{eq:fscore}) against the ground truth annotations.~\citet{leveau2004methodology}~have addressed the discrepancy for three annotation experiment participants using a small subset of the RWC music database (Goto et al., \citeyear{goto2002rwc}),\footnote{\url{https://staff.aist.go.jp/m.goto/RWC-MDB/}} including two solo bowed string instrument recordings. The authors compute the number of consistent onsets identified across the group of annotators and report the mean timing differences between consistent onsets to measure the annotation difficulty for individual recordings. 

\subsection{Motivation}\label{sec:motivation}
Standard practices for generating onset annotation ground truth typically rely on limited human input. Methodologies often involve either a single annotator whose work is validated by a small group of experts (Fonseca et al., \citeyear{fonseca2021use}), or a small set of independent expert annotations \citep{mirex2021onset}. While economical, these approaches represent a small set of views associated with the localisation of markers crucial for training and evaluation of automatic systems. The annotation of soft onsets is particularly challenging, as it is more difficult to localise start points within a time-domain signal without obvious and consistent signal characteristics for note events. It is therefore deemed important to consider multiple perspectives and strategies within the annotation process to identify problematic scenarios for both humans and algorithmic annotators, and to ensure the best annotations are chosen. However, with the exception of \citet{leveau2004methodology}, there is limited research on the agreement between multiple onset annotations for the same musical material. This makes it challenging to establish standardised guidelines for using approaches such as computation of consistent onsets to retain only the annotations of the best annotator---the one whose annotation timings most closely align with the average consistent onsets.

In this work, we conduct three experiments with 24 human annotators to investigate inter-annotator agreement, the effect of onset type (e.g., bow, finger), and the similarity between human and automated annotations. At present, onset detection datasets contain only a few string instrument examples; further, no datasets contain multiple recordings of the same piece with repeated musical instances and contexts. To mitigate this problem, we present the \textit{Virtuoso Strings} dataset which consists of audio recordings of individual string instruments performed by a professional string quartet in a variety of musical and expressive conditions. A subset of the dataset has been annotated by 24 human annotators with varying levels of musical training. We adapted the method proposed by \citet{leveau2004methodology} to select the most consistent set of annotations, aiming to create ground truth data for the training and evaluation of data-driven onset detection algorithms.

The remainder of this paper is structured as follows: Section \ref{sec:haydn_dataset} introduces the \textit{Virtuoso Strings} dataset. Section \ref{sec:exp-1} examines inter-annotator agreement and consistency, and Section \ref{sec:exp-2} investigates annotation difficulty for different note onset types and presents our approach to identifying the ground truth annotations among many annotators. Section \ref{sec:exp-3} compares the performance of five established onset detection algorithms against that of the human annotators and the ground truth annotations. Finally, Section \ref{sec:conclusions} provides concluding remarks and suggestions for future work. 

\section{Virtuoso Strings Dataset}\label{sec:haydn_dataset}
The \textit{Virtuoso Strings} dataset is a large-scale multitrack collection of string ensemble recordings designed for timing analysis, onset detection, and automatic music transcription research \citep{tomczak2023virtuoso}. It contains recordings of quartet, trio, duet, and solo performances, each captured across multiple repeated takes and under three controlled performance conditions (normal, speed-varied, and deadpan). The full dataset comprises 746 instrument tracks drawn from five musical excerpts spanning Haydn, Beethoven, Dohnányi, and Mozart.

\begin{figure}[!t]
    \centering
    \BeginAccSupp{method=plain, 
    Alt={From left to right the first violin, second violin, viola and cello performers are sitting approximately one metre apart in a concert-style quartet arrangement.}}
    \includegraphics[width=\columnwidth]{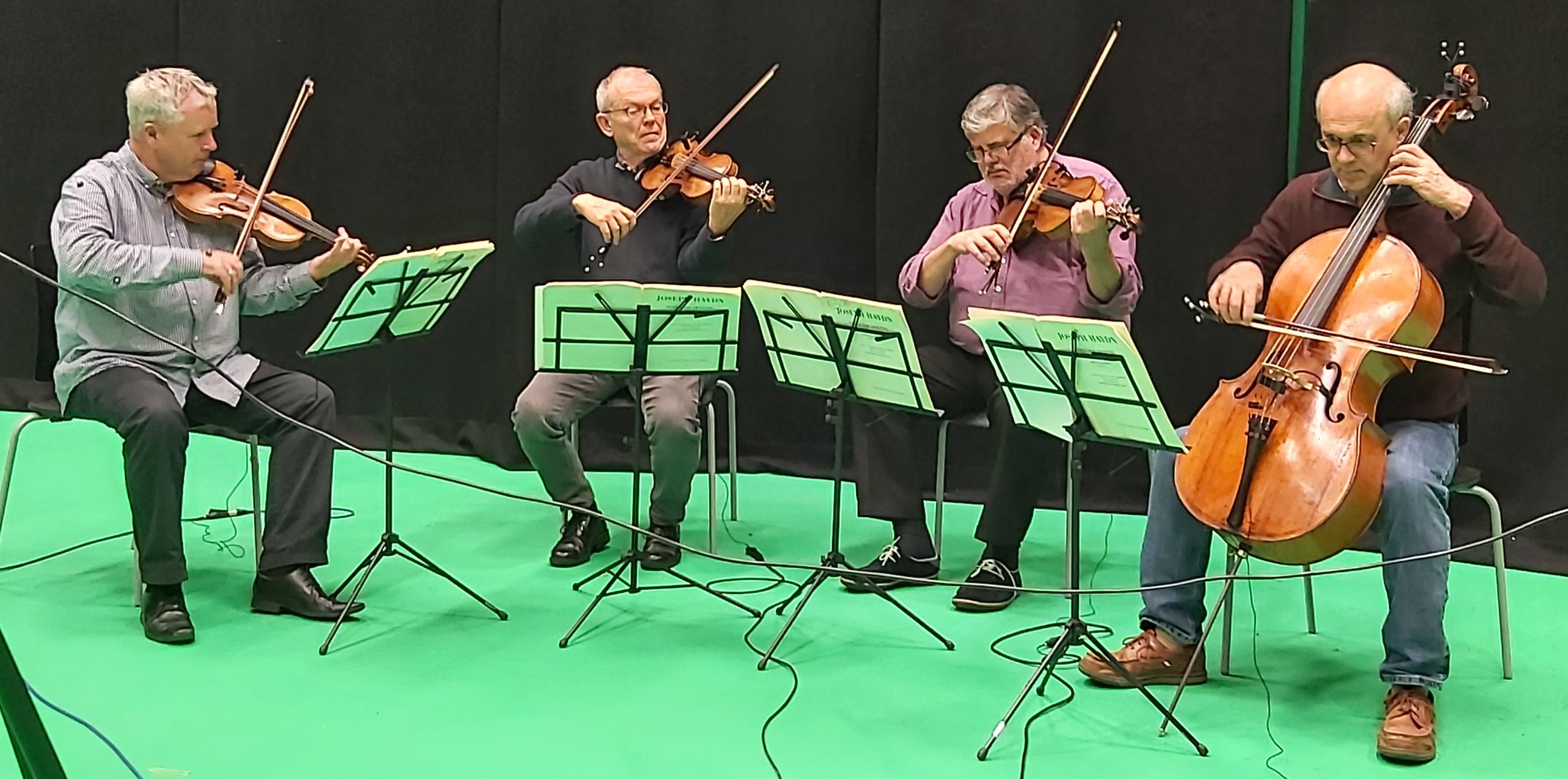}
    \caption{The Coull Quartet during a recording session.}
    \label{fig:quaret}
    \EndAccSupp{}
\end{figure}

In the present study, we make use of only a specific subset of \textit{Virtuoso Strings}, namely, recordings of the first 49 bars from the Finale of Haydn's String Quartet Op.~74 No.~1. This particular excerpt was chosen because it offers a dense mixture of soft onset types, rapid note transitions, and diverse articulation patterns, making it particularly suitable for investigating onset localisation challenges in bowed-string performances. The subset also contains twelve repeated takes under three controlled performance conditions, providing the repeated musical contexts necessary for analysing inter-annotator consistency and the behaviour of automatic onset detection algorithms. This subset includes multitrack recordings captured as part of the ARME project and is released openly for research purposes.\footnote{\url{https://github.com/arme-project/virtuoso-strings}} The corresponding scores are also available online.\footnote{\label{web:had}\url{https://github.com/arme-project/haydn-annotation-dataset}} As the present study focuses on a single Haydn excerpt performed by one string quartet, the repertoire and performer diversity considered here is necessarily limited; future work should examine whether the observed annotation patterns generalise across a broader range of composers, musical periods, and performers.

As seen from left to right in Figure \ref{fig:quaret}, the first violin~(VN1), second violin~(VN2), viola~(VA) and cello~(VC) performers sat approximately one metre apart in a concert-style quartet arrangement. Recordings were realised through the use of instrument-mounted DPA 4099 series instrument microphones, which minimised bleed between instruments and ensured high-quality isolation of each sound source. All audio recordings were captured in a mono WAV format with a 44.1\;kHz sample rate and 16-bit depth. 

The selected Haydn excerpt was recorded 12 times (i.e., in 12 separate takes) each for~3 different musical conditions resulting in 48 tracks per condition, and 144 tracks in total. The conditions represent different interpretation instructions, and were chosen to span a wide range of performance types. It is thus possible to create more intricate contexts for the annotation of soft onset types which might have otherwise not been found if only standard concert performances were analysed. The normal condition~(\normal) represents a concert-style performance, where musicians were asked to play as if on a concert stage. The speed condition~(\speed) includes spontaneous accelerando and decelerando initiated by a single musician (i.e., VN1, the leader) throughout each performance. The other musicians (i.e., VN2, VA, VC) were asked to follow the leader to the best of their ability. The leader was free to choose the magnitude of temporal variation of a single accelerando followed by a decelerando in each performance. To assess challenging contexts for onset annotation and automated detection, we included the deadpan (\deadpan) condition, for which musicians were asked to play with minimal expression in tempo and articulation. Both human and automated detection performance may then be compared against those of the \normal\;and \speed\;conditions to understand whether discrepancies arise from ambiguities in the written score or from interpretative choices made by the musicians during their performance. Additionally, as the dataset consists of multitrack files, custom instrument combinations (e.g., first violin + cello) can be easily generated.

\section{Experiment 1: Inter-annotator Agreement}\label{sec:exp-1}
In this experiment, we analyse the agreement between different annotators, reporting on the consistency and asynchrony of annotations. We first introduce the annotation participants and experimental methodology, and subsequently outline two techniques utilised in the evaluation of human annotator performance: an existing visualisation method for identifying inter-annotator agreement (Ren et al., \citeyear{ren2018investigating}), and a method for computation of average consistent onsets by \citet{leveau2004methodology}, which we extend to be generalisable to many annotators with varied musical experience.

\subsection{Participants}\label{sec:participants}

Twenty-four students and researchers from the University of Birmingham, Birmingham City University and University of Warwick took part in the experiment. Their musical experience, as defined by years of regular practice of a musical instrument and/or singing voice, ranged from no musical experience to professional musical career (minimum 0 years, maximum 48 years, mean 11.4 years). All participants completed a questionnaire prior to undertaking the experiment, which requested information about their musical and annotation experience. Participants with musical experience then provided the number of years of experience, and listed their primary musical instrument. Participants were also asked if they had any known hearing impairments. Questionnaire queries and responses are presented in Table~\ref{tab:quest} (Appendix~\ref{app:quest}) and are available on the supporting website.\footnoteref{web:had} Informed consent was obtained from all participants who contributed to this study.

\subsection{Stimuli}
Each of the 24 participants annotated the four recordings (i.e., VA, VC, VN1, VN2) of take 12 from the \normal\; condition. The annotations from this experiment are denoted as \nr12 and resulted in a total of 96 annotations, 24 overlapping per instrument. The final take (i.e., \nr12) was chosen to ensure optimal performance consistency, ensuring that the musicians had fully acclimatised to the specific experimental constraints and recording environment over the course of the preceding takes.

\subsection{Onset Annotation}
All participants were instructed to use the Sonic Visualiser~(SV) software\footnote{\url{https://www.sonicvisualiser.org/}} to annotate note onsets in each instrument's recordings following a brief training session. The visual cues were presented to each participant as waveform representations (e.g.,~top~of~Figure~\ref{fig:wav-annotations}), which provided visual orientation during event segmentation. Although annotation based only on visual cues can allow for semi-precise segmentation of some notes, every participant was instructed to confirm each annotation by listening to the recordings. Annotators were free to listen to each recording as many times as needed and to use the time-stretching functionality of SV for the more challenging sections of the performances; however, they were instructed not to use additional visual cues provided by SV (e.g., pitch tracks, spectral representations) to prevent overpowering auditory stimuli with too many visual cues. 

The resulting annotations are represented by a sequence of time points demarcating note onsets, as annotated by each participant. The number of onsets per annotator may vary due to missed or mistakenly placed onsets. To address potential human errors, all onsets within $30$\,ms were combined into a single one located at the arithmetic mean of the positions following the approach by B{\"o}ck et al. (\citeyear{bock2012evaluating}). The same preprocessing step was implemented in all experiments. 

\subsection{Method}
\subsubsection{Inter-annotator Agreement}\label{sec:onset-evaluation}
Onset detection performance is typically evaluated against a single, definitive set of ground-truth annotations provided by experts. However, in this experiment, our objective is to quantify the variability and consensus among the participant pool prior to establishing a final reference. We focus on assessing inter-annotator agreement without designating a specific participant as the ground truth. We follow a similar approach to that of Ren et al. (\citeyear{ren2018investigating}) and Toma{\v{s}}evi{\'c} et al. (\citeyear{tomavsevic2021exploring}), who represented pairwise agreement between annotators using the standard F-measure metric. 
Because precision and recall depend on which annotation set is treated as the reference, we follow \citet{tomavsevic2021exploring} and evaluate each annotator pair bidirectionally. Given two sets of annotations from different participants~(e.g.,~annotator~$i$ and annotator~$j$), we first calculate the metric treating annotator~$i$ as the reference and annotator~$j$ as the candidate, and subsequently swap roles to treat annotator~$j$ as the reference and annotator~$i$ as the candidate. The resulting pairwise scores are visualised as an agreement matrix.

The calculation of the F-measure needed in the assessment of inter-annotator agreement uses several key concepts from the task of automatic onset detection. If an event is detected in an acceptable range, it is labelled as a true positive~(TP). If a detected onset is not within this range, it is labelled as a false positive~(FP). Alternatively, if an annotated ground truth event does not coincide with any detected onset, it is labelled as a false negative~(FN). 
Based on these classifications, precision~(P) and recall~(R) can be calculated. Precision denotes the fraction of true positives among all detected events, while recall represents the fraction of true positives among the total number of ground truth events. These metrics are defined as follows:
\begin{equation}
    P = \frac{TP}{TP+FP},
\end{equation}
\begin{equation}
    R = \frac{TP}{TP+FN}.
\end{equation}
\noindent
The precision and recall can be expressed as a single metric using the F-measure (F) calculation:
\begin{equation}
    F = 2\frac{PR}{P+R}.
    \label{eq:fscore}
\end{equation}
A tolerance window is used to determine whether a detected onset is within an acceptable range on either side of the ground truth. The window is represented as a duration, for example, $\pm$25\,ms. In automatic transcription tasks, these windows are often informed by the shortest rhythmic interval within a metrical grid (e.g., a 32nd note) to avoid merging distinct events \citep{wu2018review}. 
However, musicological studies of microtiming, such as the analysis of Maracatu rhythm by \citet{fonseca2021use}, require significantly greater temporal precision to capture intentional deviations from the grid. These studies often employ stricter tolerance thresholds (e.g., $<$20\,ms) compared to the standard 20--50\,ms windows typically found in the general onset detection literature~(B{\"o}ck et al., \citeyear{bock2012evaluating}; Wu et al., \citeyear{wu2018review}).
Soft onsets in bowed-string performance can, however, exhibit perceptual ambiguity, with listeners distributing onset judgements across a temporal interval \citep{danielsen2019beat}. While adaptive or wider onset bins might better reflect this perceptual spread, they risk reducing the precision required to evaluate algorithmic latency. 
In this study, we opted for a $\pm$25\,ms evaluation window, which is a stricter criterion, as it offers more discerning results, particularly for soft onsets and online onset detection~\citep{bock2012evaluating}, compared to a broader $\pm$50\,ms window~(e.g.,~used for percussion onsets~\citep{bello2005tutorial,dixon2006onset}). 

Inter-annotator agreement is evaluated using a similarity matrix of F-measure scores which represent higher agreements for comparisons where a larger proportion of onsets were within the set tolerance window~$\omega$~($\omega$=25\,ms), a size commonly used in related work (Eyben et al., \citeyear{eyben2010universal}; B{\"o}ck et al., \citeyear{bock2012evaluating}; Ali-MacLachlan et al., \citeyear{ali2016note}).

\subsubsection{Consistent Onsets}\label{sec:consistent-onsets}
To produce a generalisable measure of annotation variance, we extend the approach from~\citet{leveau2004methodology} for determining consistent onsets. Consistent onsets are defined as coincident timestamps identified through pairwise comparisons between annotators. 
From these comparisons, we derive two metrics: average consistent onsets~(ACO), which captures the mean count of matching onsets found across comparisons, and average timing difference~(ATD), which captures the mean temporal deviation between these matching onsets. 
Higher ATD values indicate larger disagreement in how participants annotated a particular instrument. These metrics are dependent on the tolerance window size ($\omega$). The window acts as a parameter for granularity: narrower windows emphasise precision (capturing fine-grained disagreements), while broader windows reflect generalised consensus. 
In \citet{leveau2004methodology}, consensus was determined via a sequential intersection strategy (e.g., comparing subject~1~to~2, then the result to~3). This approach is order-dependent. While a pairwise approach removes this sequential bias, it introduces a challenge regarding ambiguous pairings: if multiple onsets from annotator $j$ fall within the tolerance window of a single onset from annotator $i$, the pairing choice affects the resulting timing difference. To resolve this ambiguity without bias, we employ a randomised matching process repeated $k$ times. In each repetition, ambiguous matches are resolved stochastically. We then average the results to ensure convergence. 

Let $O_i$ and $O_j$ denote the sets of onsets for annotators $i$ and $j$. The total number of unique annotator pairs is $|A| = \binom{N}{2}$. In each repetition $k$, we identify a set of matching onsets $C_{ij}^{(k)}$ for every pair $(i,j)$ within tolerance $\omega$. The timing difference $\Delta_k$ for a single repetition $k$ is calculated as the mean absolute deviation across all matching pairs that contain at least one match:
\begin{equation}
\Delta_k = \frac{1}{|A^{'}|} \sum_{(i,j) \in A^{'}} \left( \frac{1}{|C_{ij}^{(k)}|} \sum_{(t_a, t_b) \in C_{ij}^{(k)}} |t_a - t_b| \right), 
\label{eq:atd} 
\end{equation}
where $A^{'}$ denotes the subset of pairs with non-empty matches in repetition~$k$, and~$(t_a, t_b)$ represents a matched pair of timestamps. The final ATD is defined as the mean of $\Delta_k$ across all repetitions. The simulation terminates when this cumulative average stabilises (i.e., when the absolute difference between the cumulative average at repetition $k$ and repetition $k-1$ is less than 1\,ms).

We evaluate these metrics using tolerance windows ranging from 25\,ms to 100\,ms. Preliminary testing indicated that consistent onsets were difficult to identify at strict tolerances when including novice annotators. Therefore, to ensure robust identification of consistent onsets, we focus on a subset of $N=16$ annotators who possess five or more years of musical experience. As shown in Sections \ref{sec:exp-1-results} and \ref{sec:exp3-results}, this experience level is associated with higher annotation consistency.

\begin{figure}[!t]
    \centering
    \BeginAccSupp{method=plain, 
    Alt={Four boxes positioned in two rows representing pairwise F-measure scores calculated between 24 annotation participants.}}
    \includegraphics[scale=0.25]{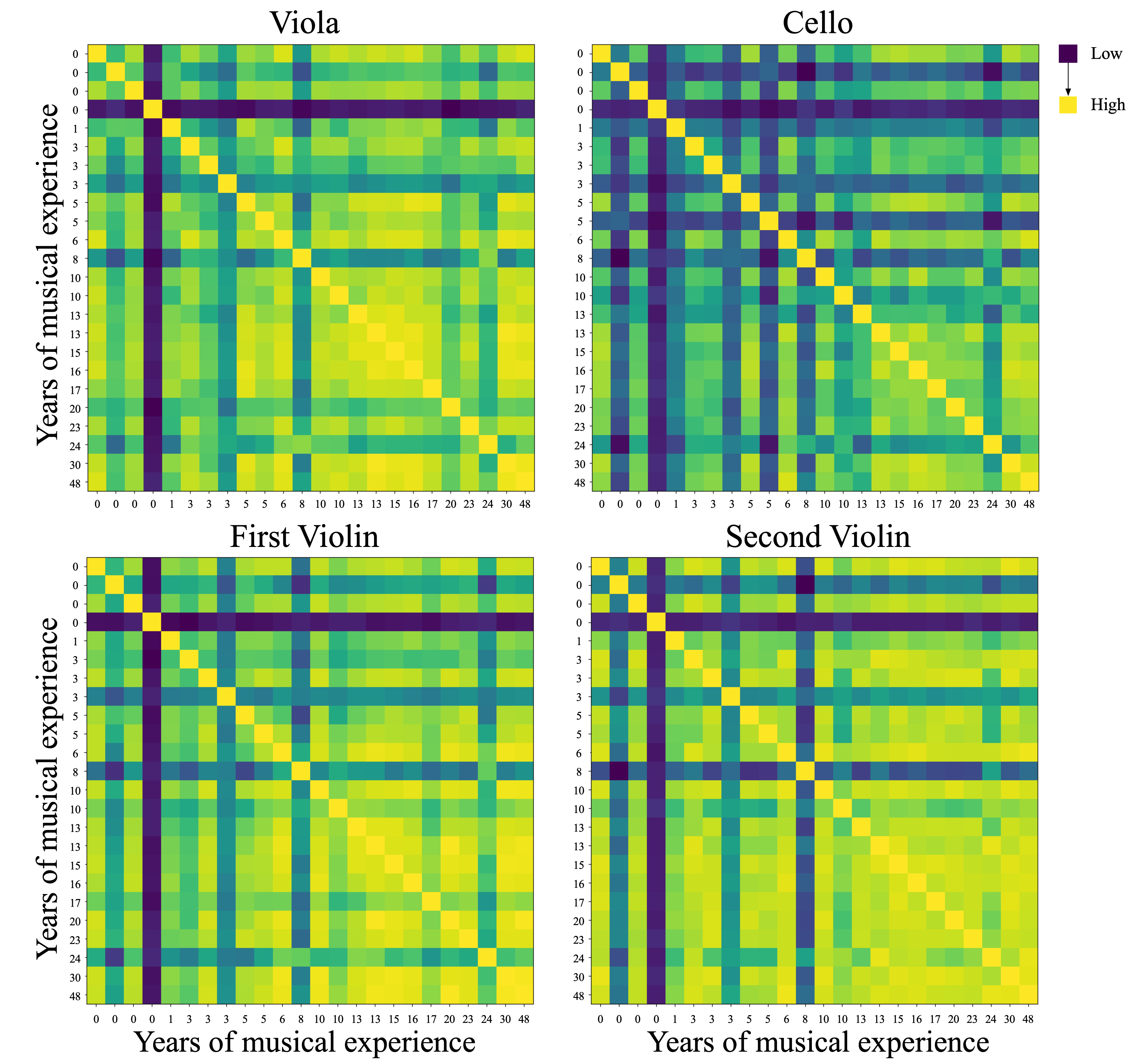}
    \caption{Pairwise F-measure scores calculated between 24 participants using a fixed tolerance window of 25\,ms. Participants are sorted from least to most years of musical experience (i.e., top to bottom and left to right). High and low scores are indicated by bright and dark colours, respectively.}
    \label{fig:f-pairwise}
\EndAccSupp{}
\end{figure}

\subsection{Results and Discussion}\label{sec:exp-1-results}
Figure \ref{fig:f-pairwise} shows matrices with pairwise F-measure scores for different instruments in~\nr12 performances, using a tolerance window of 25\,ms. The labels on the x and y axes show years of musical experience of each of the 24 annotators. The comparisons between the annotators are sorted according to the number of years of musical experience, from the least experienced to most experienced (i.e., top to bottom and left to right). 
The results highlight higher F-measures in comparisons involving annotators with longer musical experience (e.g., lower right corner of the matrices). The matrices reveal a gradient of increasing agreement with years of experience, with a noticeable decline in consistency for participants with fewer than five years of experience (top-left quadrant).
This lower agreement is particularly evident in the cello annotations, which display consistently lower F-measure scores (indicated by darker matrix cells) across the majority of pairs compared to the viola and violins. This disparity indicates a higher difficulty in the annotation of cello recordings. A similar relationship can be observed in results for tolerance windows larger than~25\,ms which are presented on the supporting website.\footnote{\label{web:demo}\url{https://arme-project.co.uk/demos/onset-annotation}}

\begin{figure}[!t]
    \centering
    \BeginAccSupp{method=plain, 
    Alt={Two subfigures with consistent onset counts and their timing differences over tolerance windows from 25 milliseconds to 100 milliseconds.}}
    \includegraphics[scale=0.48]{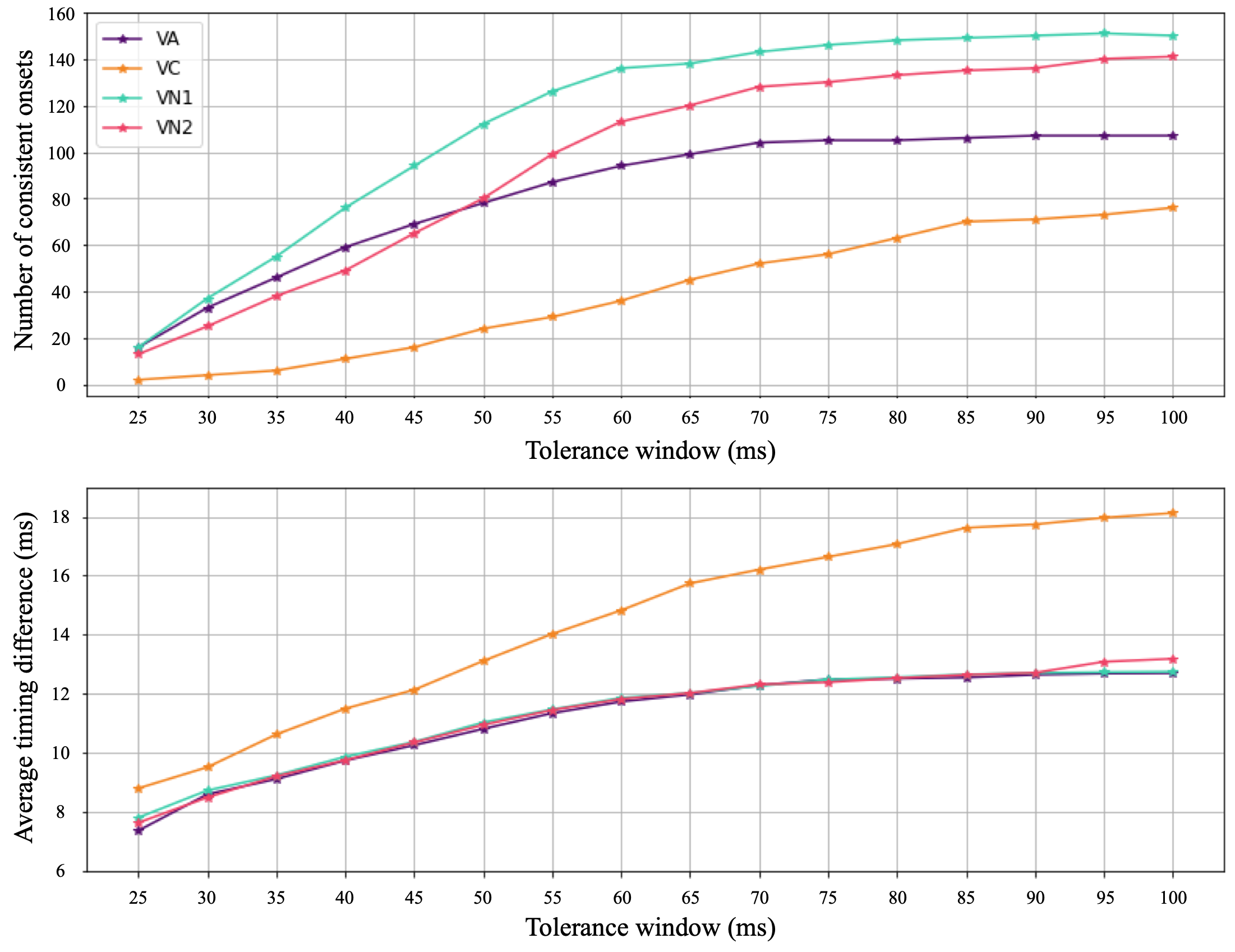}
    \caption{Number of consistent onsets (top) and average timing differences between consistent onsets~(bottom) are plotted for tolerance windows ranging from 25\,ms to 100\,ms.}
    \label{fig:consist-onsets}
\EndAccSupp{}
\end{figure}

The top panel of Figure~\ref{fig:consist-onsets} shows the number of consistent onsets from \nr12 performances, plotted for tolerance windows ranging from 25\,ms to 100\,ms, using the methodology described in Section~\ref{sec:consistent-onsets}. The total numbers of note onsets (i.e., as defined in the score) in VA, VC, VN1 and VN2 recordings are 116, 100, 167 and 150, respectively. While most consistent onset annotation results approach the total note counts at higher tolerance windows, relatively few are found at tolerances~$<$50\,ms (e.g.,~$<$20 consistent onsets for each instrument at 25\,ms). The bottom panel of Figure~\ref{fig:consist-onsets} shows average timing differences calculated from consistent onsets for different tolerance windows. Timing differences for cello annotations are larger than for viola and violins. While at smaller tolerance windows the disparity in ATD values between instruments remains relatively small (i.e., $\approx$1.5\,ms), this gap widens to approximately 5\,ms at the 100\,ms tolerance. Additionally, a slight decrease in the number of consistent onsets can be observed at the 100\,ms tolerance window for~VN1, which is due to multiple onsets overlapping, as VN1 has the shortest mean inter-onset interval~(IOI) in the~\nr12 performance, at approximately 70\,ms.

These results, as well as the pairwise F-measure results, indicate higher difficulty in annotation of cello recordings by the experiment participants. 
While the disparity in average timing differences between the cello and other instruments is small (i.e., $<$2\,ms gap at 25\,ms tolerance), it is important to note that the number of consistent onsets found for the lower tolerance window sizes is very small when compared with the total numbers of notes in the performance (see Table \ref{tab:nr-12-onsets}). For example, at the 25\,ms tolerance for VN1 there are approximately 20 consistently annotated onsets with an approximate 8\,ms average timing difference between them out of the total of 167 note onsets in the performance. This result indicates that, on average, the experiment participants did not agree with the annotation of approximately 147 note onsets within the first violin recording in \nr12. Overall, small numbers of consistent onsets for the smaller tolerance windows indicate substantial disagreement between annotators. While a general analysis of consistent onsets and average timing differences does not highlight which onsets manifest the highest annotation disagreements, we conduct a more granular investigation in the following experiment, considering different types of onsets described in Section \ref{sec:note_types}.

\section{Experiment 2: Onset Type Analysis}\label{sec:exp-2}
This experiment investigates the performance of all annotators on different note onset types in \nr12 performances. Additionally, a new set of expert annotations is created in order to help evaluate automatic approaches to the detection of note onsets, as detailed in the evaluation presented in Experiment 3 (Section \ref{sec:exp-3}). These annotations include a high level of granularity with labels for each individual note played by each musician. In addition to note labels, every note onset time also includes a separate annotation denoting onset type. Two pairs of complementary types of onsets are presented: change of bow or finger within a bow, and fingered (stopped) or open string (unstopped).

\subsection{Participants and Expert Reference Annotation}\label{sec:expert} 
While Experiment 1 yielded twenty-four unique sets of annotations, the application of standard onset detection metrics (see Section \ref{sec:onset-evaluation}) necessitates a discrete reference point. Treating the collective raw annotations as a ground truth introduces significant variance, complicating the use of such metrics. Consequently, rather than introducing an arbitrary external expert, we derived a representative reference annotation directly from the participant data through a three-stage process. This process began with centroid selection, aimed at identifying the participant whose annotations best represented the group consensus. By calculating the ACO with a tolerance of 25\,ms, we determined that Participant~2~($a_2$) exhibited the highest consistency with the group average. Subsequently, the annotations from~$a_2$ underwent verification and correction, where they were manually cross-referenced against the musical score by the authors (each possessing a minimum of 10 years of experience with string instrumentation). Crucially, this step was strictly limited to rectifying objective errors, such as missing notes or obvious outliers, thereby preserving the micro-timing characteristics of $a_2$'s original perception. The resulting verified set is designated as the expert reference annotation~($a_0$). This approach balances the reality of perceptual onset ambiguity with the methodological requirement for a precise reference. By grounding $a_0$ in the participant closest to the ACO centroid, the reference reflects the collective perception of the listeners while providing the clean, discrete data necessary for algorithmic evaluation. The $a_0$ annotations are included in the open dataset.\footnoteref{web:had}

\begin{table}[!t]
    \centering
    \BeginAccSupp{method=plain, 
    Alt={Table with note counts per onset category in viola, cello, and first and second violins.}}
    \small
    \tabcolsep=0.15cm
    \begin{tabular}{c|c|cc|cc}
    Inst. & All notes & Open string & Stopped note & Bow start & Finger change \\ \hline
    VA & 116 & 20 & 96 & 99 & 17 \\
    VC & 100 & 6 & 94 & 89 & 11 \\
    VN1 & 167 & 5 & 162 & 112 & 55 \\
    VN2 & 150 & 18 & 132 & 115 & 35
    \end{tabular}
    \caption{Number of notes per onset category in viola (VA), cello (VC), and first and second violins (VN1 and VN2) in \nr12.}
    \label{tab:nr-12-onsets}
\EndAccSupp{}
\end{table}

\subsection{Stimuli and Onset Type Annotation}
To facilitate a granular analysis of detection performance, we generated a new set of metadata for the expert annotations ($a_0$). By aligning the verified $a_0$ timestamps with the corresponding musical score—which contains specific performance instructions such as bowing markings and slurs---we were able to assign a specific note type label to each onset. 
The four onset types are categorised into two pairs; each pair comprises complementary types of onset. The first pair defines them as being either an \textit{open string} or \textit{stopped note} (i.e., any fingered note). The second pair describes onsets as a \textit{bow start} (i.e., first note within a bow direction as indicated by the score) or \textit{finger change} (i.e., any subsequent fingered notes played within a shared bow direction). Table~\ref{tab:nr-12-onsets} presents the numbers of onsets in each category.

\subsection{Results and Discussion}\label{sec:exp2-onset-type-results}
\begin{figure}[!t]
    \centering
    \BeginAccSupp{method=plain, 
    Alt={Radar chart with mean true positive rates from all annotators for four note onset types. Finger change onset type annotations for cello show lower performance in comparison with other types.}}
    \includegraphics[scale=0.5]{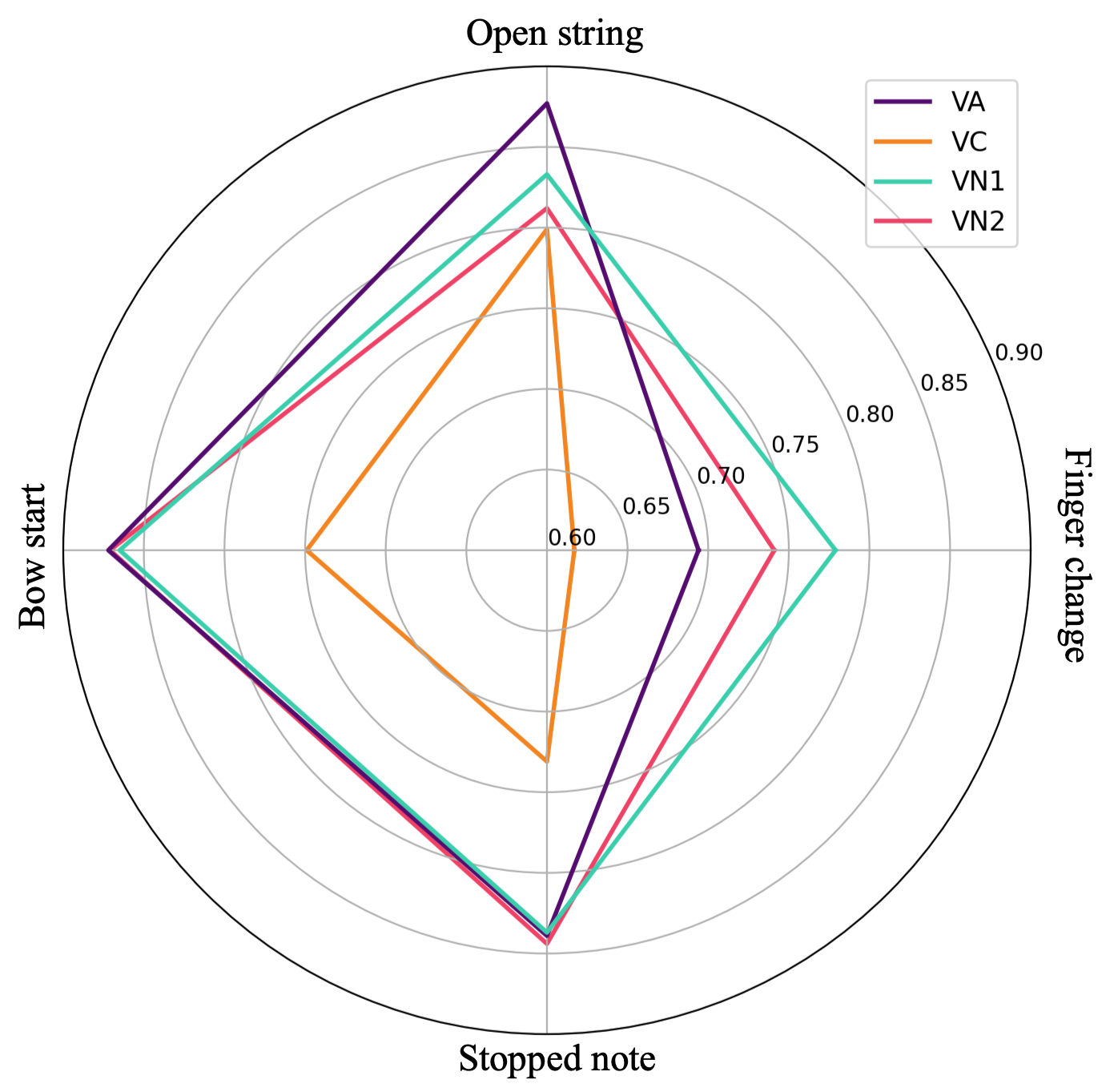}
    \caption{Mean true positive rates from all annotators calculated from comparisons against the expert reference annotation $a_0$ in \nr12.}
    \label{fig:ons-types-radar}
\EndAccSupp{}
\end{figure}

Figure \ref{fig:ons-types-radar} shows the mean performance of all annotators in relation to the expert reference annotation $a_0$ (here considered ground truth) for each instrument and onset category calculated as a percentage of TP onsets. The reported results use a tolerance window of 25\,ms and are presented as means from all annotation participants, while the per-annotator results are available on the supporting website.\footnoteref{web:demo} 

The highest mean performance from the annotators is observed in viola annotations for the open string onset category, indicating the highest agreement between all annotators for this onset type. The lowest level of agreement between all annotators is present in the cello annotations of the finger-change onset type, which also shows low agreement for the other instruments. Mean true positive rates across all participants for the open string and stopped notes as well as bow start and finger change onsets are 83.0\%, 81.3\%, 83.9\% and 70.8\%, respectively. Additionally, the mean true positive rates across all participants for VA, VC, VN1 and VN2 are 82.0\%, 72.4\%, 82.8\% and 81.7\%, respectively. Overall, given the relatively small tolerance window there seems to be a stronger annotation agreement for all onset types except those of finger change onsets, which can be predominantly characterised through pitch change rather than modification of transient information.

Regarding the lower annotation agreement for the cello, especially concerning finger changes, several factors may contribute. The physical properties of the instrument provide one plausible explanation; cello strings possess significantly higher mass and inertia compared to violin strings \citep{rossing1998physics}. Guettler (\citeyear{guettler2006violin, guettler2002bowed}) suggests that the buildup time (i.e., the time required for steady-state oscillation to develop) is inversely proportional to frequency. Consequently, the cello may exhibit a longer physical onset duration than the violin or viola, creating a less defined transient region~\citep{akar2020investigation}. However, given that the present study involves a single quartet and cellist, performer-, musical-part-, and recording-specific effects cannot be excluded. 
Finally, finger changes constitute soft onsets defined primarily by pitch shifts rather than energy transients, a challenge further compounded by performance variables such as string tension, tuning, and player idiosyncrasies. To address these complexities, future work should explore models that exploit pitch-related and longer-term temporal information \citep{bittner2017deep,tomczak2023onset}. However, the success of these models relies heavily on the availability of domain-specific data; evaluations in \citet{stefani2024importance} highlight that even small amounts of instrument-specific training data can improve detection performance, underscoring the need for dedicated resources and data augmentation strategies where real-world examples are scarce.

\section{Experiment 3: Onset Detection}\label{sec:exp-3}
In this experiment, we evaluate annotations for \normal, \speed\;and \deadpan\;conditions using five established onset detection algorithms spanning signal-processing and neural-network approaches. The goal is to benchmark signal processing and deep learning systems within a common evaluation framework and to highlight their performance relative to the proposed taxonomy of soft onset types. Additionally, the highest-performing system is selected for a comparative analysis of individual human annotators. For completeness and reproducibility, extended evaluation results and all experimental data are provided on our supporting website.\footnoteref{web:demo}

\subsection{Participants}
Four participants were selected from the previous experiments to take part in a third study due to their musical experience (minimum 5 years). The participants include $a_{12}$, $a_{13}$, $a_{16}$ and $a_{18}$ with 13, 5, 10 and 16 years of musical experience, respectively. In addition to the annotations from the four participants, we also use the expert reference annotation $a_0$ for \nr12.

\subsection{Stimuli}
Participants annotated 140 recordings, totalling the remaining repetitions 1--11 for \normal\;and repetitions 1--12 for \speed\;and \deadpan\;conditions. A different number of recordings was assigned to each participant with the mean of 35 recordings per participant. As in the first experiment, the annotation experiment participants were not constrained by time and could use the time-stretching functionality of SV during the more challenging sections of each recording. Each participant was asked to use the default primary view within SV for consistency purposes. The duration each annotator spent on the task varied based on individual familiarity and pace. All participants performed their annotations for this experiment after annotating the recordings used in Experiment 1.

\subsection{Method}
\subsubsection{Onset Detection Algorithms}\label{sec:exp3-algorithms}
In this experiment, the human-generated annotations serve as the reference ground truth, and the automatic onset detection algorithms are evaluated against these human labels. The highest-performing algorithm is subsequently used only as an additional comparative benchmark to contextualise annotator performance, rather than as a source of reference annotations. 
We selected a representative suite of algorithms available in the \texttt{madmom}\footnote{\url{https://github.com/CPJKU/madmom}} Python library (B{\"o}ck et al., \citeyear{madmom}). 
This selection enables us to compare distinct and widely accessible approaches, ranging from signal processing to deep learning, within a unified framework, while using these algorithms to illuminate their capabilities and limitations relative to human perception and to provide a baseline for future work on automatic soft-onset detection. 
SuperFlux~(\SuF) by \citet{bock2013maximum} computes the difference between the short-time magnitudes of nearby spectral frames and is optimised for music signals with soft onsets and vibrato effect in string instruments. ComplexFlux~(\CoF) by \citet{bock2013local} is based on the \SuF\;algorithm with the addition of local group delay measure which makes this method more robust against loudness variations of steady tones. We also use the standard spectral flux~(\SpF) method by \citet{masri1996computer}. To compute detection functions from the three methods, we use log-filtered spectrograms as inputs with default parameters provided by \texttt{madmom}. 
Additionally, to compute activation functions, we use the pretrained recurrent neural network~(\RNN) approach of~B{\"o}ck~et~al.~(\citeyear{bock2012online}) and the convolutional neural network~(\CNN) approach of \citet{schluter2014improved}, as distributed with \texttt{madmom} v0.16.1; no training or fine-tuning was performed in this study. Specifically, we use the default \texttt{RNNOnsetProcessor} with \texttt{online=False}, corresponding to the bidirectional RNN configuration, and the default \texttt{CNNOnsetProcessor}. Of the evaluated methods, the CNN approach of \citet{schluter2014improved} is the most recent.

\subsubsection{Peak-Picking Strategy}
Once a detection or an activation function $\theta$ has been generated for an instrument recording under observation, the onsets must be located in the temporal domain from~$\theta$. We adapt the thresholding approach of Eyben et al. (\citeyear{eyben2010universal}) for onset detection. A threshold $T$ can be determined using the mean of all frames in $\theta$ and a constant $\lambda$:
\begin{equation}
T = \operatorname{mean}(\theta)\lambda.
\end{equation}
If the current frame $\eta$ is a local maximum above $T$, then the onset function $\Gamma$ is given by:
\begin{equation}
    \Gamma(\eta) = 
    \begin{cases}
    1, &\theta(\eta-1) \:\:<\:\: \theta(\eta) \:\:\geq\:\:  \theta(\eta+1) \:\:\&\:\: T < \theta(\eta) \\
    0, &\textit{otherwise.}
    \end{cases}
\end{equation}

\subsubsection{Evaluation Metrics}
Standard precision (P), recall (R), and F-measure~(F) scores (see Section \ref{sec:onset-evaluation}) are used for evaluation of the onset detection performance. 
Consistent with the methodology in Experiment 1, we utilise a fixed tolerance window of $\pm$25\,ms. This stricter criterion (compared to the standard 50\,ms) is chosen to rigorously evaluate temporal precision against the reference annotations, distinguishing accurate detections from loose approximations. 
The mean F, P and R results are calculated as averages across annotations from all participants per instrument and onset detection method. The~$\lambda$ parameter is chosen using grid search across the \nr12 dataset and is optimised for highest F-measure, with optimisation performed once per detection method across all four instruments.

\begin{figure*}[!t]
    \centering
    \BeginAccSupp{method=plain, 
    Alt={Five section bar chart for five automatic systems with F-measure, precision, recall scores plotted for NR12, NR, SP and DP conditions.}}
    \includegraphics[scale=0.43]{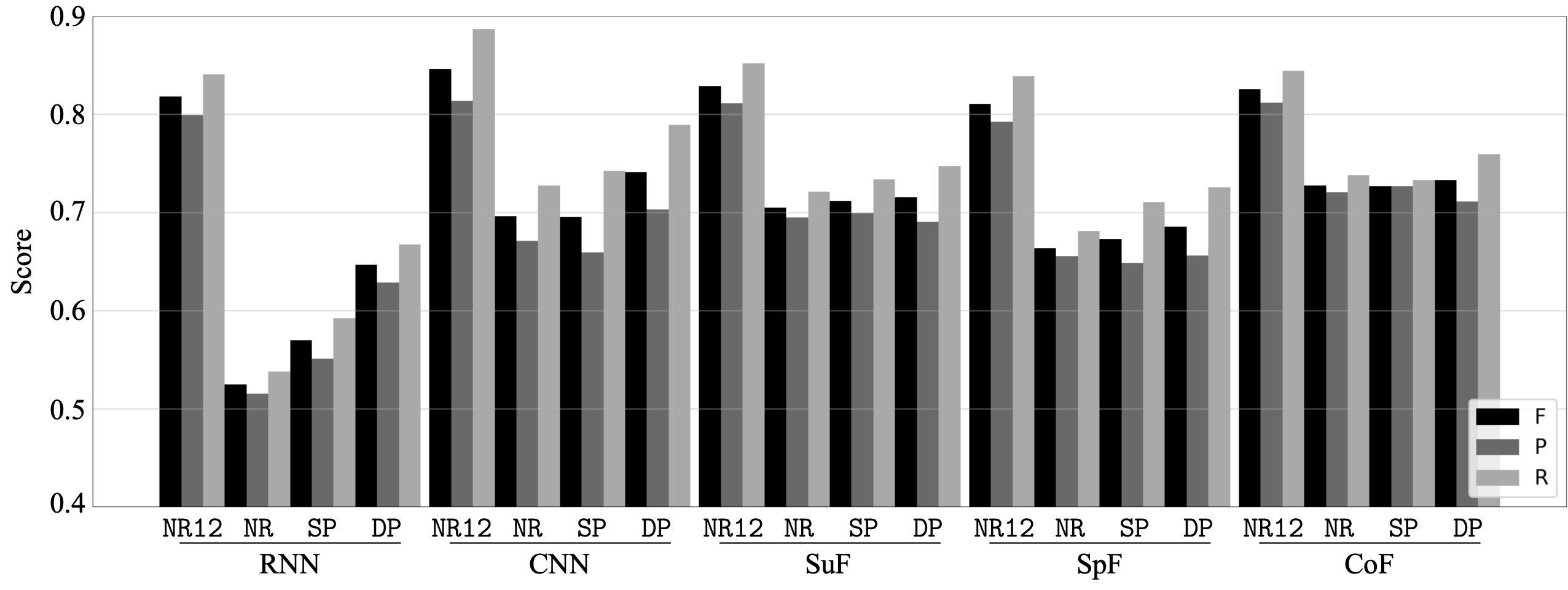}
    \caption{Mean F-measure (F), precision (P) and recall (R) calculated across VA, VC, VN1 and VN2. Different onset detection methods are compared against the expert reference annotation $a_0$ in \nr12 and annotations from $a_{12}$, $a_{13}$, $a_{16}$ and $a_{18}$ in the \normal, \speed\;and \deadpan\;conditions. Results for \nr12 represent the calibration set, while results for \normal, \speed\;and \deadpan\;represent evaluation on the remaining 140 recordings.}
    \label{fig:od-conditions}
\EndAccSupp{}
\end{figure*}

\subsection{Results and Discussion}\label{sec:exp3-results}
Figure \ref{fig:od-conditions} presents the correspondence between algorithm-detected onsets and human annotations. Results for \nr12 should be interpreted as calibration-set performance because the detection thresholds were optimised on these four recordings, whereas results for the \normal, \speed\;and \deadpan\;conditions provide evaluation on the remaining 140 recordings. 
The highest-performing algorithms across all instruments are \CNN\;and \CoF. While all methods perform similarly on the \nr12 recordings, a noticeable performance drop occurs for the \normal, \speed\;and \deadpan\;conditions. This is likely due to differences in annotation style between the expert reference annotation $a_0$ (see Section \ref{sec:expert}) and the participant group ($a_{12,13,16,18}$), as well as the increased musical complexity of the larger dataset.
The \RNN\;method exhibits the largest performance drop, suggesting a higher sensitivity to this variability. 
Specifically, the \speed~condition involves rapid \textit{accelerandi} resulting in short inter-onset intervals, which increases the likelihood of the previous note's decay masking the subsequent attack. Conversely, the \deadpan~condition minimises articulatory accents, resulting in less defined transients that are acoustically harder to distinguish from the steady-state signal. 
This musical heterogeneity can lead to larger fluctuations in algorithm performance when averaged across many performances, unlike the more uniform and limited set of four \nr12 recordings. The \normal, \speed, and \deadpan~conditions include 44, 48, and 48 recordings, respectively, naturally encompassing a wider range of expressive and interpretative differences than the tightly controlled \nr12~set. While mean precision scores are similar across all algorithms on recordings with \nr12{} expert annotations, the highest mean precision (0.814) and recall (0.887) are achieved by the \CNN{} system.

\begin{table}[!t]
    \centering
    \BeginAccSupp{method=plain, 
    Alt={Table with true positive rates for different onset types, instruments and automatic onset detection systems. Finger change onsets show lower mean detection performance than the other onset types across all algorithms.}}
    \small
    \tabcolsep=0.1cm
    \begin{tabular}{c|c|cc|cc}
    Inst. & Method & Open string & Stopped note & Bow start & Finger change \\ \hline
    VA &  & 20 onsets & 96 onsets & 99 onsets & 17 onsets \\ \hline
     & \RNN & 0.85 & 0.83 & 0.87 & 0.65 \\
     & \CNN & 0.95 & 0.90 & 0.94 & 0.71 \\
    VA & \SuF & 1.00 & 0.86 & 0.97 & 0.41 \\
     & \SpF & 0.95 & 0.85 & 0.89 & 0.76 \\
     & \CoF & 1.00 & 0.85 & 0.97 & 0.35 \\ \hline
    VC &  & 6 onsets & 94 onsets & 89 onsets & 11 onsets \\ \hline
     & \RNN & 1.00 & 0.86 & 0.89 & 0.73 \\
     & \CNN & 1.00 & 0.81 & 0.83 & 0.73 \\
    VC & \SuF & 1.00 & 0.87 & 0.94 & 0.36 \\
     & \SpF & 1.00 & 0.85 & 0.90 & 0.55 \\
     & \CoF & 1.00 & 0.86 & 0.93 & 0.36 \\ \hline
    VN1 &  & 5 onsets & 162 onsets & 112 onsets & 55 onsets \\ \hline
     & \RNN & 0.80 & 0.79 & 0.74 & 0.89 \\
     & \CNN & 1.00 & 0.91 & 0.94 & 0.85 \\
    VN1 & \SuF & 1.00 & 0.81 & 0.91 & 0.64 \\
     & \SpF & 1.00 & 0.79 & 0.78 & 0.84 \\
     & \CoF & 1.00 & 0.80 & 0.88 & 0.65 \\ \hline
    VN2 &  & 18 onsets & 132 onsets & 115 onsets & 35 onsets \\ \hline
     & \RNN & 0.89 & 0.86 & 0.89 & 0.80 \\
     & \CNN & 0.89 & 0.92 & 0.97 & 0.74 \\
    VN2 & \SuF & 0.89 & 0.81 & 0.93 & 0.46 \\
     & \SpF & 0.94 & 0.81 & 0.90 & 0.57 \\
     & \CoF & 0.89 & 0.82 & 0.94 & 0.46 \\ \hline
     & \RNN & 0.88 & 0.84 & 0.85 & 0.77 \\
     & \CNN & 0.96 & 0.88 & 0.92 & 0.76 \\
    Means & \SuF & 0.97 & 0.84 & 0.94 & 0.47 \\
     & \SpF & 0.97 & 0.83 & 0.87 & 0.68 \\
     & \CoF & 0.97 & 0.83 & 0.93 & 0.46 
    \end{tabular}
    \caption{Calibration-set true positive rates for different onset types using the expert reference annotation $a_0$ for \nr12.}
    \label{tab:ons-accuracies}
\EndAccSupp{}
\end{table}

Onset detection results for different note onset types using the expert reference annotation $a_0$ for \nr12 are presented in Table \ref{tab:ons-accuracies}. The results are reported as true positive rates~(i.e., recall) per onset type, instrument, and algorithm. Mean results for each algorithm across instruments are shown at the bottom of the table. The highest-performing method is the \CNN\;system with a mean recall of 0.879 across all onset types. While the results indicate overall high detection rates for open string, stopped note and bow start onsets, the finger change onsets display much lower recall scores. 
The subtle and low-energy nature of finger-change onsets provides a plausible explanation for their lower detection performance. Unlike open-string or bow-start onsets, which generate clear transient energy and pronounced spectral changes, finger-change events often involve gradual pitch transitions with only minimal modification to the sound envelope. These cues are easily masked by ongoing resonance, particularly in lower instruments \citep{fletcher2012physics}, and are therefore difficult both for human annotators to perceive and for onset detection algorithms to capture reliably, since most algorithms rely on abrupt temporal or spectral discontinuities. This hypothesis is consistent with the findings from Experiment~2 (Section~\ref{sec:exp2-onset-type-results}), in which finger-change onsets also yielded the lowest annotation agreement among participants, underscoring the inherent ambiguity of this onset type.
\begin{figure}[!t]
    \centering
    \BeginAccSupp{method=plain, 
    Alt={Three bar charts for NR, SP and DP conditions showing F-measure scores plotted for viola, cello, first violin and second violin annotations from annotators 12, 13, 16 and 18. Annotator 18 has the highest mean F-measure.}}
    \includegraphics[scale=0.45]{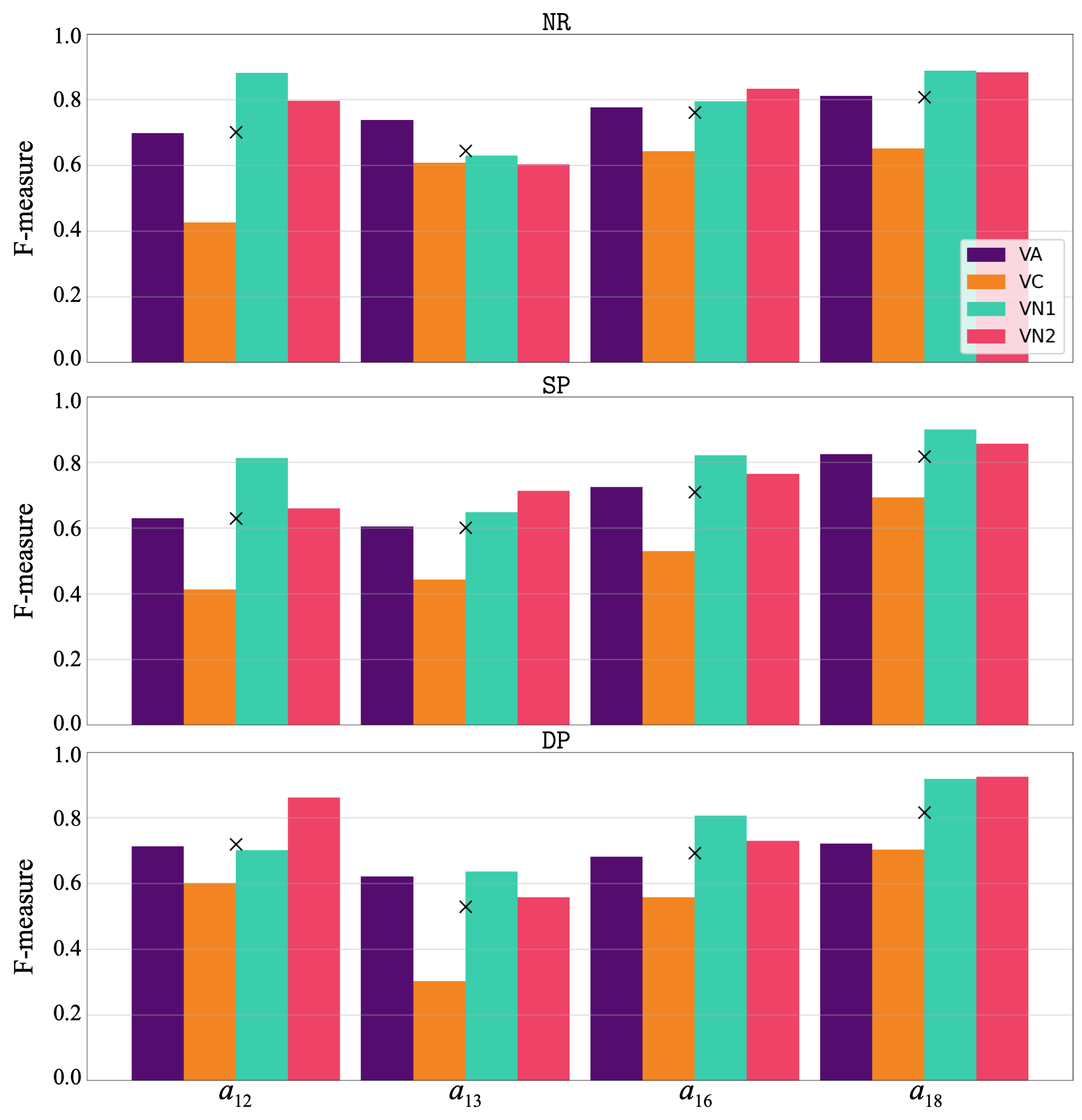}
    \caption{Mean F-measures acquired using the highest-performing \CNN\;onset detection system for \normal\;(top), \speed\;(centre) and \deadpan\;(bottom) conditions. The results are reported for annotators $a_{12}$, $a_{13}$, $a_{16}$ and $a_{18}$ with per annotator means ($\times$).}
    \label{fig:od-cnn}
\EndAccSupp{}
\end{figure}
Interestingly, the overall high-performing \CNN\;and \RNN\;systems show lower results for the viola and cello bow start onsets when compared to the \SuF\;and \CoF\;methods but outperform these methods in the detection of finger change onsets. The lower performance of \SuF\;and \CoF\;for the finger change onsets can also be seen in other instruments. This indicates that the deep learning systems demonstrate a more uniformly high performance over all note onset types while the signal processing methods struggle with the more challenging finger change onsets. Additionally, the more easily detected open string onsets (compared with stopped notes) might be partially explained by the fact that the engagement of an open string often marks a change of string (e.g., from D to A string) and is thus associated with the fresh bow attack, even within a bow stroke, involved in such a move. A similar ease of identification of the open string note onsets can be observed in Experiment 2, where the annotators show high agreement (see Figure \ref{fig:ons-types-radar}) in the annotation of this type of onset across all instruments. 

We also report on mean F-measures for each annotator who participated in this experiment. Figure \ref{fig:od-cnn} presents results obtained using the highest-performing~\CNN\;system, in which the detections are compared against onsets (i.e., ground truth) from each of the four participants. The~\CNN\;system achieved the highest means~(shown as~$\times$) across instruments for annotator~$a_{18}$ in all conditions; 0.81 in~\normal, 0.82 in~\speed\;and 0.82 in~\deadpan. Additionally, the lowest mean detection performance is observed for the cello~(0.55), and the highest for the first violin~(0.79). 
For annotator $a_{18}$, the cello results in the~\deadpan\;condition are similar to those for the viola, which are lower than in the~\normal\;and~\speed\;conditions. Lower viola results in~\deadpan\;are also observed for annotators $a_{13}$ and $a_{16}$. 
Moreover, the higher difficulty in accurately localising cello note onsets may be influenced by multiple factors. From an automatic detection perspective, the cello's lower frequency range introduces a time-frequency resolution trade-off inherent to the short-time Fourier transform~(STFT). Accurately resolving lower fundamental frequencies ($f_0$) requires longer analysis windows, which reduce temporal precision and may smear the onset detection function in the time domain \citep{muller2011signal}. The cello's physical properties may also contribute to less clearly defined attack characteristics, while sustained resonance can mask the subtle temporal cues associated with soft onsets \citep{fletcher2012physics}. These factors may therefore contribute to the lower performance observed for the cello. Its acoustic characteristics may make soft onsets challenging for both human annotators and onset detection algorithms, while the STFT trade-off introduces an additional limitation for automatic detection.

\section{Conclusions}\label{sec:conclusions}
In this work, we presented a comprehensive study on the annotation of soft onsets, accompanied by a new dataset of 24 human annotation sets and expert ground truth for the Finale of Haydn's String Quartet Op.~74 No.~1. By analysing these annotations alongside automatic detection algorithms, we identified critical challenges in the definition and localisation of soft onsets. 

We proposed a taxonomy distinguishing between \textit{bowed}, \textit{fingered}, and \textit{plucked} onsets, which allowed us to isolate specific articulation types that cause disagreement. Our evaluation revealed that \textit{finger change} onsets are the most challenging type for both human annotators and automated systems to identify reliably. Furthermore, we found the cello to be the most difficult instrument to annotate. The instrument's physical properties, including its greater string mass and lower frequency range, provide a plausible explanation for this result; however, given that the present study involves a single quartet and cellist, performer-, musical-part-, and recording-specific effects cannot be excluded. Additionally, our results confirm that musical experience is associated with annotation consistency, informing best practices for the selection of annotators in ground-truth generation tasks.

To support reproducibility and further research, both the high-quality multitrack recordings and the corresponding annotations are released as open datasets. Future work should extend the evaluation to additional performers and repertoire from a broader range of musical periods to determine how strongly the observed instrument- and onset-type effects generalise beyond the present quartet. The \textit{Virtuoso Strings} dataset could also be leveraged to investigate differences across repeated takes and performance conditions. In addition, future onset detection systems could explicitly model the temporal uncertainty associated with soft-onset annotations and investigate instrument-specific approaches to challenging onset types. The multitrack nature of the collection provides a useful resource for developing and evaluating such approaches.

\section*{Acknowledgements}
This project was funded by the Engineering and Physical Sciences Research Council~(EPSRC) grant with reference EP/V034987/1. We would like to thank the musicians of the Coull Quartet for their dedication and professionalism, as well as all experiment participants who contributed to this study.

\bibliographystyle{apacite}
\bibliography{references}

\newpage
\appendix
\section{Questionnaire}\label{app:quest}

\begin{table}[h]
\BeginAccSupp{method=plain, 
Alt={Table with questionnaire results from 24 annotators.}}
\small
\renewcommand{\arraystretch}{0.4}
\centering
\begin{tabular}{c|c|c|c|c|c|c}
ID & Q1 & Q2 & Q3 & Q4 & Q5 & Q6 \\ \hline \hline
1 & yes & no & 17 & piano & no & 23 \\
2 & yes & no & 48 & cello & no & 55 \\
3 & yes & no & 24 & piano & no & 28 \\
4 & yes & yes & 10 & guitar & no & 36 \\
5 & no & yes & 3 & dj-ing & no & 44 \\
6 & yes & yes & 20 & piano & no & 25 \\
7 & no & no & 0 & none & no & 30 \\
8 & yes & no & 8 & clarinet & no & 25 \\
9 & no & no & 0 & none & no & 21 \\
10 & no & no & 23 & guitar & no & 37 \\
11 & no & no & 0 & none & no & 27 \\
12 & yes & yes & 13 & trumpet/cornet & no & 21 \\
13 & yes & no & 5 & none & no & 21 \\
14 & yes & yes & 13 & guitar & no & 28 \\
15 & yes & no & 3 & guitar & no & 24 \\
16 & yes & no & 10 & oboe & no & 25 \\
17 & yes & no & 3 & vocals & no & 21 \\
18 & yes & no & 16 & piano & no & 20 \\
19 & yes & yes & 15 & guitar & no & 26 \\
20 & yes & yes & 30 & piano & no & 38 \\
21 & no & yes & 1 & guitar & no & 23 \\
22 & yes & no & 5 & piano & no & 27 \\
23 & yes & no & 6 & guitar & no & 25 \\
24 & no & no & 0 & none & yes & 24
\end{tabular}
\caption{\textnormal{Questionnaire described in Section \ref{sec:participants}. Questions Q1--Q6 are as follows: Q1: Have you studied music or are you a professional musician?; Q2: Have you ever annotated musical events before (e.g., using Sonic Visualiser)?; Q3: Years of experience playing music? (e.g., 0 if none); Q4: What is your primary musical instrument? (e.g., \textit{none} if you do not play musical instruments); Q5: Do you have any known hearing impairment?; Q6: Age.}} \label{tab:quest}
\EndAccSupp{}
\end{table}

\end{document}